\documentclass[aps,prl,twocolumn,superscriptaddress,showpacs]{revtex4-1}

\usepackage{graphicx,subfigure}
\usepackage{amssymb,bm,color}
\usepackage[nice]{nicefrac}

\begin{document}

\title{Non-axisymmetric instability of shear-banded Taylor-Couette flow}

\author{Alexandre Nicolas}
\affiliation{SUPA, School of Physics \& Astronomy, University of
  Edinburgh, JCMB, King's Buildings, Mayfield Road, EH9 3JZ,
  Edinburgh, UK}
\affiliation{Ecole Polytechnique, 91128 Palaiseau, France}
\author{Alexander Morozov}
\email{Alexander.Morozov@ph.ed.ac.uk}
\affiliation{SUPA, School of Physics \& Astronomy, University of Edinburgh, JCMB, King's Buildings, Mayfield Road, EH9 3JZ, Edinburgh, UK}

\date{\today}

\begin{abstract}
Recent experiments show that shear-banded flows of semi-dilute worm-like micelles
in Taylor-Couette geometry exhibit a flow instability in the form of Taylor-like vortices. Here
we perform the non-axisymmetric linear stability analysis of the diffusive
Johnson-Segalman model of shear banding and show that the nature of
this instability depends on the applied shear rate. For the
experimentally relevant parameters, we find that at the beginning
of the stress plateau the instability is driven by the interface
between the bands, while most of the stress plateau is occupied by the
bulk instability of the high-shear-rate band. Our work significantly
alters the recently proposed stability diagram of shear-banded flows
based on axisymmetric analysis.
\end{abstract}


\maketitle

Flows of complex fluids often differ dramatically from their Newtonian
counterparts \cite{Larson:book}. One of the most striking examples of such differences is
the phenomenon of shear-banding which is widespread in flows of
micellar solutions \cite{Larson:book,Cates:2006,*Lerouge:2010}, granular media \cite{Hecke:2010}, foams \cite{Gilbreth:2006} and
colloidal glasses \cite{Chen:1992,*Rut:2010}. Its name is derived from the observation that
when sheared these materials often split into two or more bands of
different shear rates, viscosities and viscoelastic properties
\cite{Larson:book,Cates:2006,Lerouge:2010}. In semi-dilute
solutions of worm-like micelles, shear-banding is often associated
with the so-called \emph{shear-stress plateau} -- the range of applied
shear rates $\dot\gamma_{app}$ for which the shear stress $\Sigma$ is
approximately independent of
$\dot\gamma_{app}$ \cite{Rehage:1991,Cates:1996,Cates:2006}.
Outside the stress plateau, simple 1D shear flow remains homogeneous, 
while for the values of $\dot\gamma_{app}$ on the plateau,
it splits into two regions of simple
shear flow with different shear rates \cite{Cates:2006}.

Recent experiments with semi-dilute worm-like micellar solutions have demonstrated that this picture of steady 1D
shear-banded flows only holds \emph{on average} 
\cite{Lerouge:2006,Becu:2007,Lerouge:2008,Fardin:2009}. 
Visualisation experiments in flows between two rotating coaxial
cylinders, the Taylor-Couette flow, have shown that on the
stress plateau, there exist large
instantaneous fluctuations in local stresses and the local position of the
interface between the two bands that are associated with a 3D
hydrodynamic instability
\cite{Lerouge:2006,Becu:2007,Lerouge:2008}. 
This instability has the form of doughnut-shaped vortices threaded
  by the inner cylinder and stacked in
the vorticity direction (Taylor-like vortices) and is mainly
localised in the high-shear-rate band \cite{Fardin:2009}. Its origin
is not inertial due to the low flow velocities and large viscosities
involved \cite{Cates:2006,*Lerouge:2010}.

Presently, the precise nature of the instability driving the Taylor-like vortices and
undulations of the interface remains unclear. 
Possible
mechanisms include an instability driven by the presence of the
interface (\emph{interfacial mode}) or a bulk instability inside one or
both bands (\emph{bulk mode}).  The interfacial mode is a viscoelastic
analogue of the Kelvin-Helmholtz instability and arises in co-flows of
several liquids with a discontinuity in their normal-stress differences 
across the interface \cite{Renardy:1988,*Hinch:1992,*Renardy:1999}. 
The bulk mode, which has been observed in
Taylor-Couette flow of polymer solutions, is driven by large values of
the first normal stress difference $N_1$ in the bulk, and the presence
of curved streamlines \cite{Larson:1990,Pakdel:1996}.
Unlike Newtonian Taylor-Couette flow where the first instability
results in an axisymmetric stack of Taylor vortices, linear stability
analysis predicts that viscoelastic Taylor-Couette flow is unstable 
towards non-axisymmetric (wavy in the azimuthal direction) Taylor-like vortices [16].

Previous attempts to uncover the nature of the instability in the
shear-banded Taylor-Couette flow focused on the \emph{axisymmetric}
version of this flow \cite{Fielding:2010}. By performing numerical
simulations, Fielding has concluded that micellar solutions split into
three groups depending on their material properties:
fluids with low values of the first normal stress difference 
$N_1$ in their high-shear-rate band exhibit the interfacial instability
\cite{Fielding:2005,*Fielding:2006,*Wilson:2006,*Fielding:2007a},
while fluids with high values of $N_1$ exhibit the bulk instability
in curved geometries \cite{Fielding:2010}. For intermediate values
of $N_1$, the flow was predicted to be stable, and the precise position
of this region was found to depend on the curvature of the
Taylor-Couette cell \cite{Fielding:2010}. The main problem with this
scenario is that the fluid used by Lerouge 
\emph{et al.} \cite{Lerouge:2006,Lerouge:2008,Fardin:2009} to observe
the instability most probably belongs to the \emph{stable} region of the stability diagram proposed
in \cite{Fielding:2010} (see Supplemental Material), 
casting doubt on the proposed mechanism.

In this Letter we show that this contradiction is resolved if one
relaxes the assumption of axisymmetric flow. We perform a linear
stability analysis of the shear-banded Taylor-Couette flow with
respect to non-axisymmetric disturbances for a model fluid with 
material properties similar to the solutions used in the experiments by Lerouge 
\emph{et al.} \cite{Lerouge:2006,Lerouge:2008,Fardin:2009}. We find
that while axisymmetric perturbations are stable as predicted by
Fielding \cite{Fielding:2010}, non-axisymmetric modes are unstable for
\emph{all} applied shear rates on the stress plateau. The nature of the
instability depends on the applied shear rate: very close to
the beginning of the plateau, we find this is dominated by
the interfacial mode, while most of the plateau is
unstable via the bulk mode in agreement with the observations by Fardin
\emph{et al.} \cite{Fardin:2009}.

Our model consists of the momentum conservation equation in the limit
of negligible inertia,
\begin{equation}
-\bm\nabla p + \eta_s \Delta \bm v  + \nabla\cdot \bm\tau = 0,
\label{ns}
\end{equation}
and the diffusive Johnson-Segalman (DJS) equation 
\cite{Larson:book,Johnson:1977,Olmsted:2000,Fielding:2005,Fielding:2010}
for the viscoelastic stress $\bm\tau$:
\begin{eqnarray}
&&\bm\tau + \lambda \left(
  \frac{1+a}{2}\stackrel{\triangledown}{\bm\tau} 
+ \frac{1-a}{2}\stackrel{\vartriangle}{\bm\tau} \right) -
l^2 \nabla^2\bm\tau \nonumber \\
&& \qquad\qquad = \eta_p \left( \bm\nabla \bm v + \bm \nabla \bm
  v^\dagger\right),
\label{JS}
\end{eqnarray}
where $\stackrel{\triangledown}{\bm\tau} = \partial \bm\tau/\partial t
+ {\bm v} \cdot {\bm\nabla}{\bm\tau} - {\bm\nabla\bm v}^\dagger\cdot{\bm\tau} -
 {\bm\tau}\cdot {\bm\nabla \bm v}$ and $\stackrel{\vartriangle}{\bm\tau} = \partial \bm\tau/\partial t
+ {\bm v} \cdot {\bm\nabla}{\bm\tau} + {\bm\nabla\bm v}\cdot{\bm\tau} +
 {\bm\tau}\cdot {\bm\nabla \bm v}^\dagger$ are the upper- and
 lower-convected derivatives, respectively \cite{Larson:book};
 $(\bm\nabla\bm v)_{ij}=\partial v_j/\partial x_i$ is the velocity
 gradient tensor, and $\dagger$ denotes the transpose. Here,
 $\bm v$ and $p$ are the velocity and pressure, $\eta_s$
 and $\eta_p$ are the viscosities of the Newtonian ``solvent'' and viscoelastic (micellar)
 components, and $\lambda$ is the Maxwell relaxation time.
 The slip parameter $a$ controls the degree of non-affine
 deformations under flow \cite{Johnson:1977} and Eq.(\ref{JS}) predicts
 a non-monotonic flow curve (shear-banding) as long as $a\ne\pm 1$ and
 $\eta_p/\eta_s>8$. The
 stress-diffusion term $l^2 \nabla^2\bm\tau$ in Eq.(\ref{JS}) prevents stress variations on
 scales smaller than $l$ and uniquely selects the value of the plateau shear stress \cite{Olmsted:2000}.
The lengthscale $l$ sets the size of structural correlations in the
fluid, and controls the width of the interface between the bands and
the extent of the near-wall layers where the bulk behaviour changes 
rapidly in order to match the boundary conditions.
Finally, we assume that the fluid is
incompressible, $\bm\nabla\cdot\bm v=0$, 
and apply the no-slip boundary conditions for the velocity $\bm v$ and
the no-flux boundary conditions for the stress tensor $\bm\tau$ \cite{Fielding:2005,Fielding:2010}.

\begin{figure}[t]
\subfigure{\includegraphics[width=4.25cm]{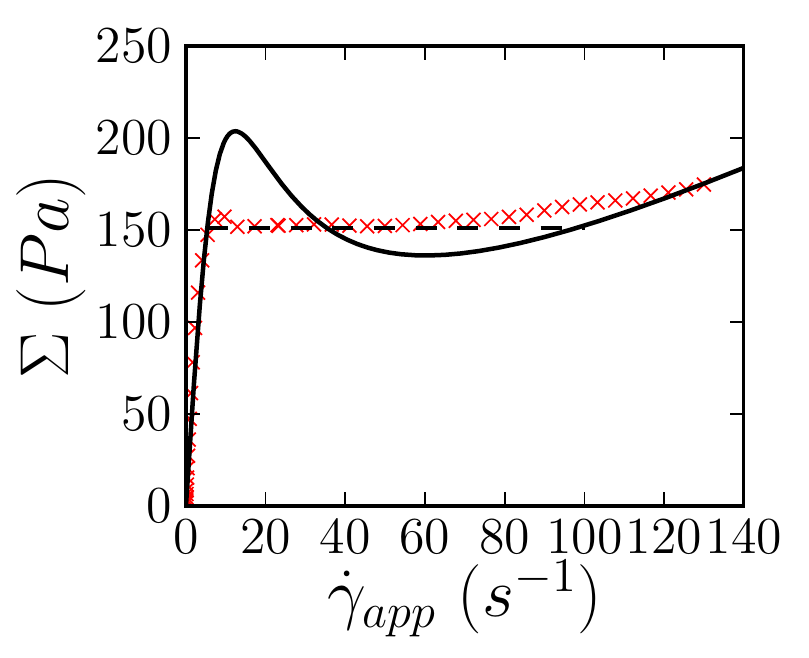}}
\subfigure{\includegraphics[width=4.25cm]{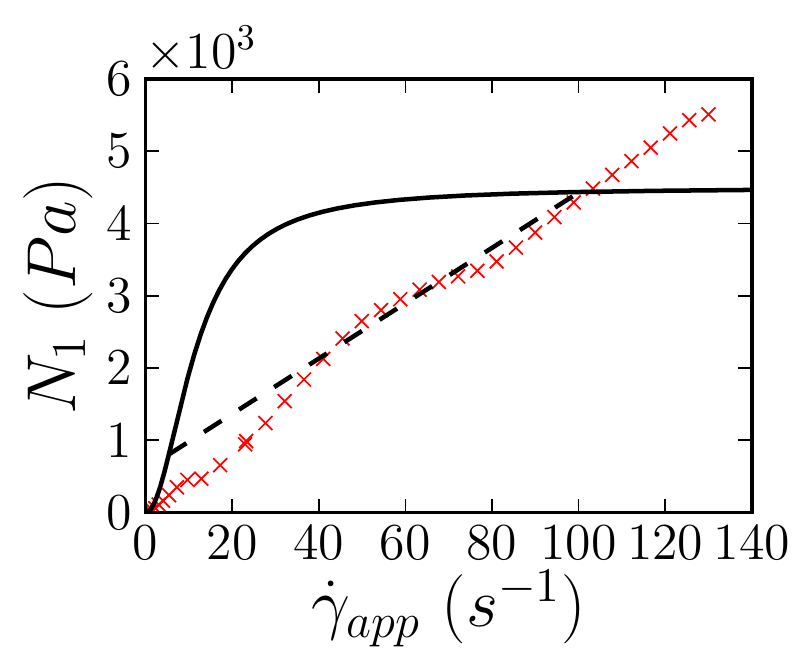}}
\caption{Total shear stress $\Sigma$ and first normal-stress difference $N_{1}$
as a function of the applied shear rate $\dot{\gamma}_{app}$. Crosses
- experimental measurements by S. Lerouge \emph{et al.}
\cite{Lerouge:2008} in a cone-and-plate geometry; solid line - JS constitutive curve for 
planar homogeneous shear flow; dashed line - prediction of the JS
model in planar shear-banded flow.}
\label{rheology}
\end{figure}

We consider the Taylor-Couette flow between two coaxial cylinders. The
inner cylinder of radius $R_1$ rotates with the angular velocity
$\Omega$, while the outer cylinder of radius $R_2$ is kept stationary.
The fluid flows in the gap of size $d=R_2-R_1$ and the relative curvature of
the Taylor-Couette cell is set by $\epsilon = d/R_1$. We choose the
parameters in our model, Eqs.(\ref{ns},\ref{JS}), to closely match
their experimental values as used by Lerouge \emph{et al.} 
\cite{Lerouge:2006,Lerouge:2008,Fardin:2009} and set $R_1=13.33\,mm$ and
$d=1.13\,mm$.  To find the relaxation time $\lambda$, the slip
parameters $a$ and the viscosities $\eta_s$ and $\eta_p$, we fit
the rheological predictions of the Johnson-Segalman model in plane
homogeneous shear flow
to the cone-and-plate measurements for a solution used in
\cite{Lerouge:2008}.

In Fig.\ref{rheology} we compare the experimental
data (crosses, courtesy of S. Lerouge) against the fit to the
Johnson-Segalman model (solid and dashed lines) with $\lambda=0.51\,s$, $a=0.985$,
$\eta_s=1.1\,Pa\cdot s$ and $\eta_p=33.0\,Pa\cdot s$. These parameters
are similar to the linear rheological measurements of
\cite{Lerouge:2008}, $\lambda\approx0.23\,s$ and $\eta_s+\eta_p\approx
55\,Pa\cdot s$. We mostly consider two
values of the structural lengthscale, 
$l=13\,\mu m$ and $l=4\,\mu m$,
which are consistent with the experimentally determined values
\cite{Ballesta:2007,*Masselon:2008,Lerouge:2008,Fardin:2010}.

\begin{figure}[t]
\subfigure{\includegraphics[width=4.25cm]{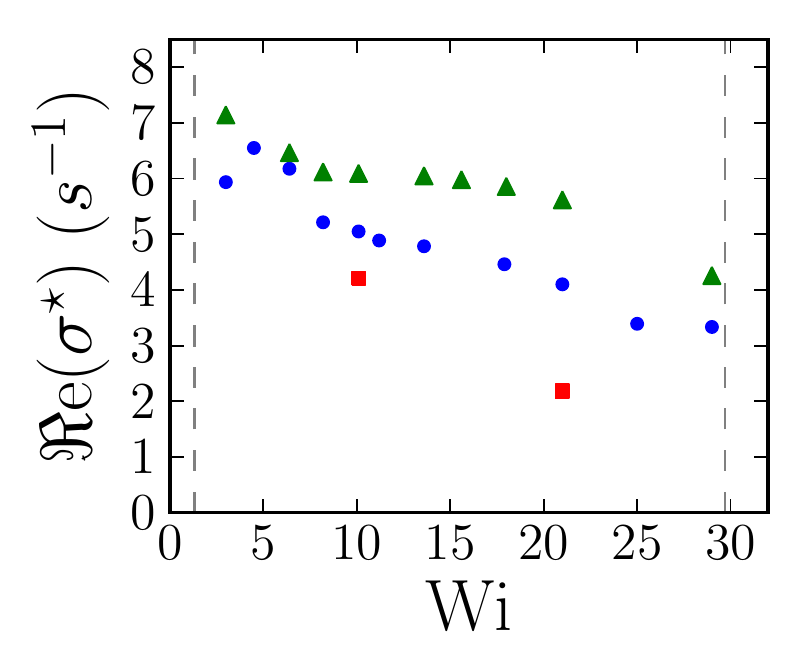}}
\subfigure{\includegraphics[width=4.25cm]{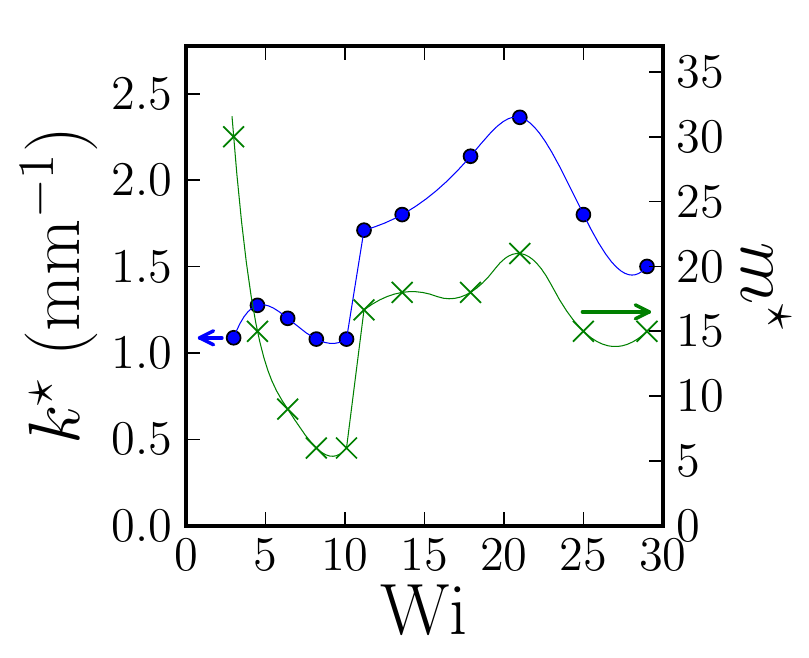}}
\caption{(left) Maximum growth rate $\Re e(\sigma^{\star})$ as a function of
Wi on the stress plateau (boundaries denoted by the dashed lines) for various
values of the diffusion length $l$: squares -- $l=73\,\mu m$,
circles -- $l=13\,\mu m$, triangles --
$l=4\,\mu m$. (right) Axial and azimuthal wavelengths $k^*$
and $m^*$ of the most unstable mode as a function of $Wi$ for
$l=13\,\mu m$. The lines are drawn to guide the eye.
}
\label{peakeig}
\end{figure}

First we introduce cylindrical coordinates $\left(r,\theta,z\right)$
and solve the equations of motion (\ref{ns},\ref{JS}) numerically
for the steady 1D shear-banded profile using a pseudospectral
Chebyshev-tau method \cite{Canuto:book}. To accurately resolve sharp
interfaces we split the computational domain into three regions, two
for each shear band and one corresponding to the interface, and use
adaptive domain decomposition algorithm to discretize them independently. A similar
method was employed in \cite{Cromer:2011}.

\begin{figure*}[t]
\subfigure{\includegraphics[width=5.5cm]{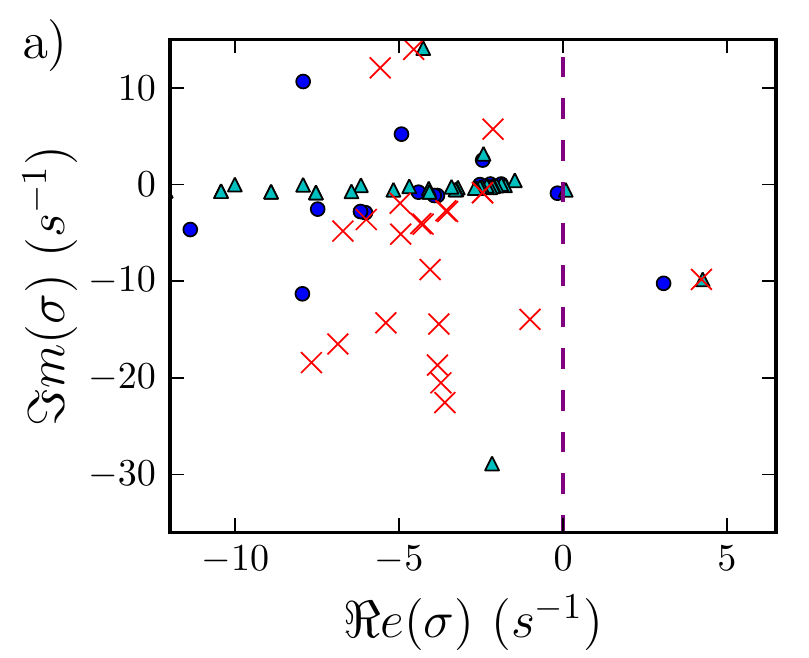}\label{highWispec}}
\subfigure{\includegraphics[width=5.5cm]{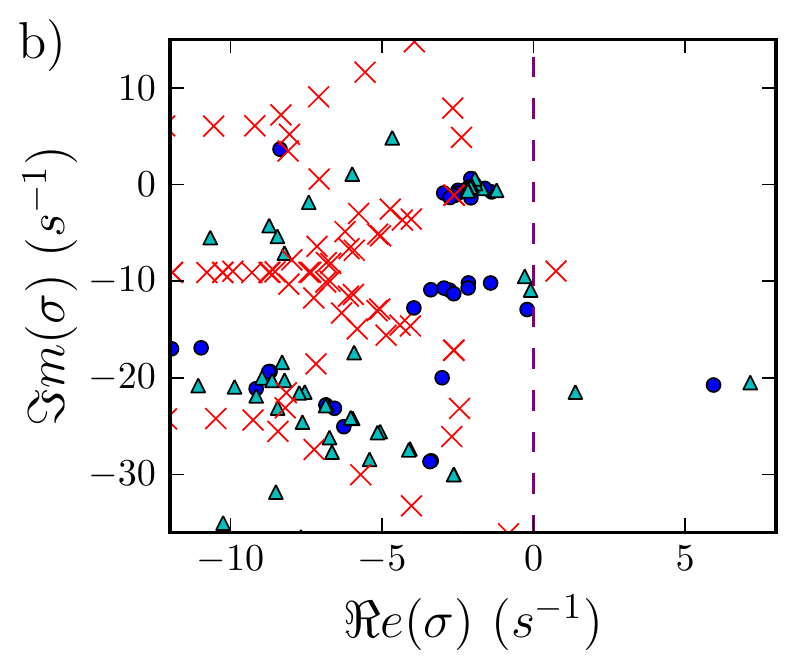}\label{lowWispec}}\\
\subfigure{\includegraphics[width=5.5cm]{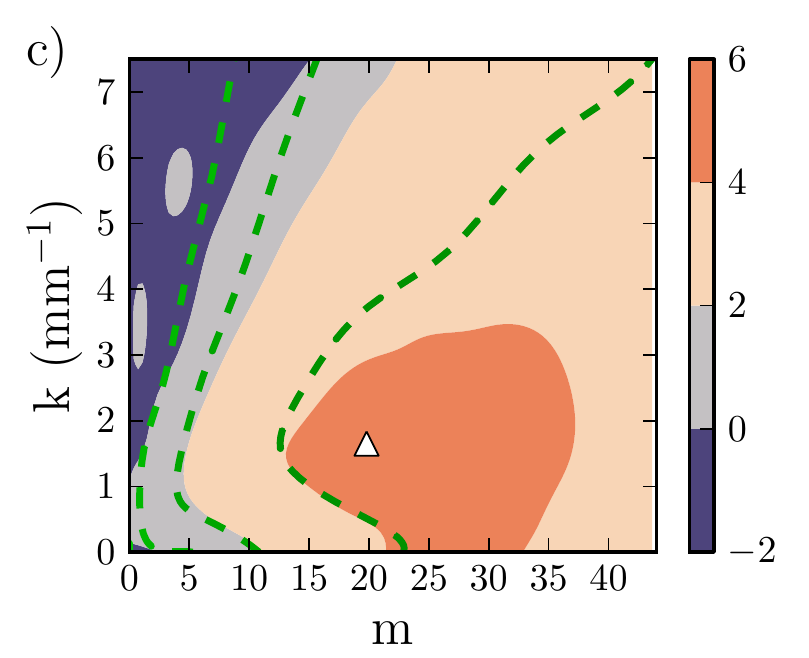}\label{highWicont}}
\subfigure{\includegraphics[width=5.5cm]{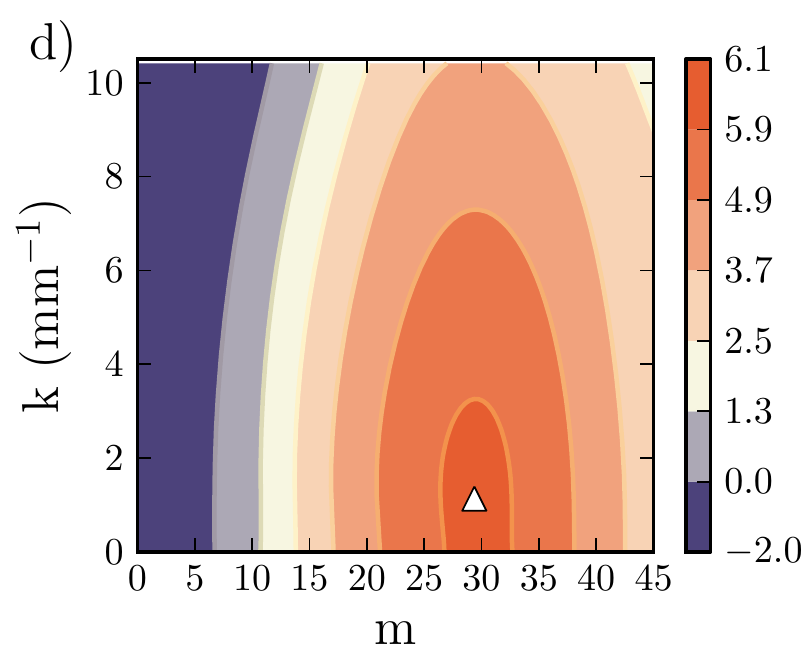}\label{lowWicont}}
\subfigure{\includegraphics[width=5.5cm]{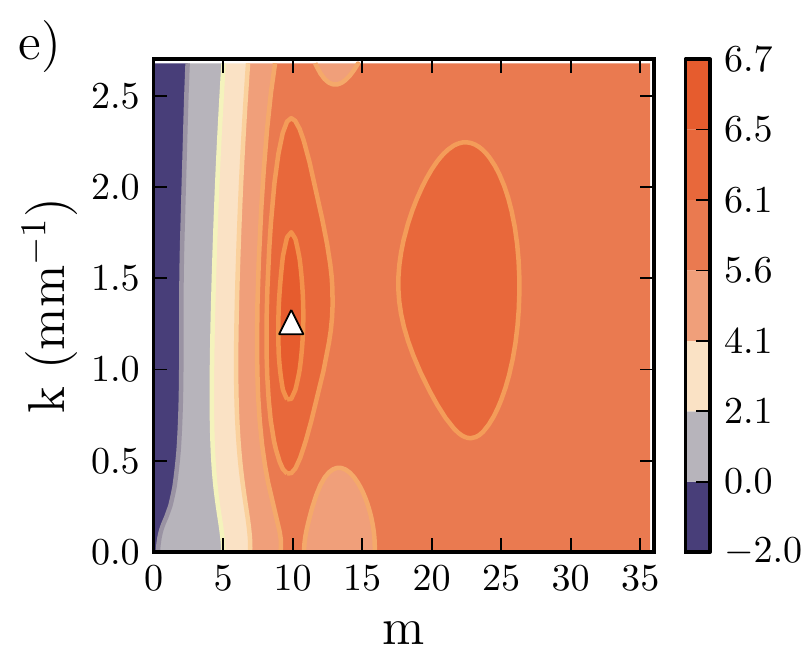}\label{intermidWi}}
\caption{(top) Superimposed spectra showing the most unstable
  eigenvalue for the shear-banded
 (dots -- $l=13\,\mu m$, triangles -- $l=4\,\mu m$) and the
 auxiliary  (crosses -- $l=4\,\mu m$) setup. (a) $Wi=29$, $k=20.54$
 and $m=21$; (b) $Wi=3.0$, $k=14.4$
  and $m=30$.
(bottom) Contour plots of the dispersion relation $\Re e\,\sigma(k,m)$
as a function of $m$ and $k$. White triangle marks
the most unstable eigenmode. (c) $Wi=29$ and
$l=4\,\mu m$, green dashed lines -- contour plot for the auxiliary system; (d) $Wi=3.0$ and $l=13\,\mu m$; (e) $Wi=6.4$ and
$l=4\,\mu m$ showing two types of instability.}
\end{figure*}

The linear stability analysis is performed by perturbing the 1D
profile with 3D perturbations of the form $(\bm v,\bm\tau,p)\sim
\exp\left(\sigma t + i k z+i m\theta\right)$, linearising the
equations of motion, and solving the resulting generalised eigenvalue
problem. Below, the applied
shear rate is reported in terms of the dimensionless
Weissenberg number $Wi=\lambda\Omega/\left(\epsilon(\epsilon+2)\right)$.

In Fig.\ref{peakeig} we plot the real part of the most unstable
eigenvalue $\sigma^*$ and the corresponding wavelengths $k^*$ and $m^*$ as a
function of $Wi$. We find that the shear-banded profile is unstable on
the whole plateau, contrary to the \emph{axisymmetric} prediction for
our parameters \cite{Fielding:2010}.

In order to determine whether the
instability is interfacial or bulk, we employ the following
method. For every $Wi$ 
along the stress plateau, we construct an auxiliary Taylor-Couette
system that models only the high-shear-rate band of the original flow.
The flow in the auxiliary system is homogeneous (not shear-banded) and
has the same laminar velocity and stress profiles as the original
high-shear-rate band. This is achieved by choosing the gap size of the
auxiliary flow cell to be
equal to the width of the high-shear-rate band and rotating both inner
and outer cylinders, while keeping $R_1$ and the DJS parameters the
same as above. 
As the result, the high-shear-rate band and its model version 
only differ in the nature of their outer boundary -- the interface with 
the low-shear- rate band (soft) or a wall (hard), respectively. (For a
recent discussion of the soft vs hard boundary conditions in the
context of shear-banding, see \cite{Fardin:2011}.)
We then compare the eigenspectra in both setups: the eigenmodes that appear in
both systems clearly correspond to the bulk instability. No
instability was found when the same procedure was applied to the
low-shear-rate band.

In Fig.\ref{highWispec} we show the comparison of the shear-banded
and (homogeneous) auxiliary systems at $Wi=29$ (near the upper end
  of the stress plateau) for $k^*$ and $m^*$, the most unstable
values of $k$ and $m$. In Fig.\ref{highWicont} we present the contour
plot of the real part of the leading eigenvalue for different values of
$k$ and $m$ in both setups. 
In spite of the different boundary conditions in the two systems, we
observe that their leading eigenvalues coincide, which we attribute to
the small size of the low-shear-rate band at high Wi and the influence
of the no-slip boundary condition at the wall on the high-shear-rate
band. This coincidence implies that the instability originates in
the bulk of the high-shear-rate band.
Moreover, we find no local maxima besides 
the eigenvalue family corresponding to the most unstable bulk mode, 
which is dominant in a wide range of $Wi\sim 10-30$. The fact that we see a 3D
instability where the axisymmetric analysis predicted stable flow
\cite{Fielding:2010}, is perhaps not surprising since the first
instability predicted for purely elastic Taylor-Couette flow of polymer solutions, which is similar to
our bulk instability, is non-axisymmetric \cite{Avgousti:1993}.

For small values of the applied shear rate the situation is
different. As can be seen from Fig.\ref{lowWispec} for $Wi=3.0$, the most unstable mode
of the shear-banded profile does not coincide with any eigenvalue of
 the auxiliary system, and we conclude that this instability is driven
 by the interface. We also observe, Fig.\ref{lowWicont}, that the
 shape of this eigenvalue family on the $\Re e\,\sigma(k,m)$ contour plot is
 different from the bulk mode of Fig.\ref{highWicont},
 and is similar to what was found by Fielding for the interfacial mode
 in plane shear
 \cite{Fielding:2005}.

At intermediate values of the applied shear rate, both the interfacial and
bulk modes are unstable, as can be seen from
Fig.\ref{intermidWi}. The velocity profiles corresponding to the 
two peaks in the $\Re e\,\sigma(k,m)$ contour plot differ fundamentally: one
has its maximum velocity in the vicinity of the interface, while the
other is mostly present in the bulk (see Supplemental Material). The height of the 
bulk peak rapidly overtakes that of the
interfacial peak as $Wi$ is increased. The transition from the
interfacial to the bulk mode takes place at the critical value of the
Weissenberg number that depends on the diffusive length:
$Wi_{crit}\approx10$ for $l=13\,\mu m$, and $Wi_{crit}\approx6-8$ for $l=4\,\mu m$.

Our use of the auxiliary Taylor-Couette
systems to determine the nature of the instability is motivated by the
observation that decreasing the value of the diffusion length $l$ 
enhances \emph{both} modes of instability. Indeed, this parameter not
only controls the width of the interface, thus affecting the
interfacial mode 
\cite{Fielding:2005,*Fielding:2006,*Wilson:2006,*Fielding:2007a},
but also sets the minimal spatial extent of stress gradients in the
fluid, and hence has a strong effect on the bulk mode as
well. Therefore, one cannot deduce the nature of an eigenmode 
by observing how its eigenvalue changes with $l$.

In summary, we presented theoretical evidence that the shear-banded
Taylor-Couette flow is unstable with respect to non-axisymmetric perturbations. For
parameters matched to the experiments by Lerouge \emph{et al.} 
\cite{Lerouge:2006,Lerouge:2008,Fardin:2009}, we find the interfacial
instability only at the beginning of the stress plateau, while most of the
plateau is occupied by the bulk instability. These results are
consistent with the observations by Fardin \emph{et al.}
\cite{Fardin:2009}, where the Taylor vortices, localised mostly in the
high-shear-rate band, were observed in a wide
range of shear rates on the stress plateau. 
Moreover, we find instabilities with the axial wavelengths of order of a few
millimeters, roughly in agreement with the asymptotic (non-linear)
wavelengths observed by Lerouge et al. [10]. Potentially, our
prediction is further supported by recent experiments of Decruppe \emph{et
  al.} \cite{Decruppe:2010}, who observed \emph{azimuthal} undulations
of the interface. However, unlike
\cite{Lerouge:2006,Lerouge:2008,Fardin:2009},  no axial interface perturbations were
found in their experiment, and its relevance to our work remains an open question.

As was noted by Fielding \cite{Fielding:2010}, the Taylor-Couette flow of worm-like
micellar solutions is unique as it brings together three types of
instabilities: shear banding itself and the interfacial and bulk
ones. The stability diagram proposed in \cite{Fielding:2010}, based on the axisymmetric
analysis, included regions of interfacial and bulk instabilities
separated by a window of stable shear-banded flow, and any given
micellar solution was predicted to belong to only one of these regions. 
Our results, based on the non-axisymmetric linear stability analysis,
suggest that both types of instabilities can be found for a given micellar
solution. We speculate that the whole curvature-$N_1$ stability
diagram may be occupied by the unstable region, with the position of
the transition between the interfacial and bulk modes, $Wi_{crit}$,
being determined by the normal stresses in the high-shear-rate band:
for larger values of $N_1$ the transition would happen at smaller $Wi_{crit}$.

\begin{acknowledgments}
We would like to thank Sandra Lerouge for providing the rheological data
used in Fig.\ref{rheology}, and Mike Cates, Marc Fardin and Sandra
Lerouge for useful discussions. AM acknowledges support from the 
EPSRC Career Acceleration Fellowship (grant EP/I004262/1).
\end{acknowledgments}

\section{Supplemental Material}
\subsection{Position of the solution used by Lerouge \emph{et al.} \cite{Lerouge:2008} on the stability diagram of Fielding \cite{Fielding:2010}}

Since the relative curvature of the flow cell used by Lerouge \emph{et al.} \cite{Lerouge:2008} was fixed at $1.13/13.33\approx 0.08$, the stability of the shear-banded flow as predicted by Fielding \cite{Fielding:2010} only depends on the model parameter $\nicefrac{1}{(1-a)}$. For this value of the curvature, the fluid was found to exhibit the interfacial instability for $\nicefrac{1}{(1-a)}<10$, the bulk instability for $\nicefrac{1}{(1-a)}>600$, or to be stable for intermediate values of $\nicefrac{1}{(1-a)}$ \cite{Fielding:2010}. The position of the boundaries between different types of instability are dependent on the parameter $\eta_s/\eta_p$, and the values quoted above are for $\eta_s/\eta_p=0.05$ \cite{Fielding:2010}.

In plane homogeneous (not banded) shear flow, the JS model predicts the following
expression for the first normal stress difference $N_1$ as
a function of the applied shear rate $\dot\gamma_{app}$:
\begin{equation}
N_1=\frac{2\lambda\eta_p\dot\gamma_{app}^2}{1+\lambda^2\dot\gamma_{app}^2\left(1-a^2\right)}.
\label{N1}
\end{equation}
Eq.(\ref{N1}) predicts that $N_1$ approaches a horizontal asymptote as $\dot\gamma_{app}$ tends to infinity,
and the normal-stress difference of the high-shear-rate band $N_1^{h}$ is well approximated by the asymptotic value of $N_1$.  In the limit  $a \rightarrow 1$, Eq.(\ref{N1})  yields:
\begin{equation}
N_1^{h} \approx \frac{\eta_p}{\lambda \left(1-a\right)},
\label{N1h}
\end{equation}
and hence, the model parameter $\nicefrac{1}{(1-a)}$ is proportional to the value of the first normal stress difference in the high-shear-rate band. Using the data of S. Lerouge, Fig.1 of the main text, we estimate $N_1^{h}\approx3-5\cdot10^3\,Pa$. The relaxation time in the linear regime was found to be $\lambda\approx 0.23s$ \cite{Lerouge:2008}, as mentioned in the main text. From the slope of the $\Sigma(\dot\gamma_{app})$ at small applied shear rates, the total shear viscosity is in the range of $20-50 Pa\cdot s$ which is consistent with the linear rheology value $55 Pa\cdot s$ found in \cite{Lerouge:2008}. Since $\eta_s\ll\eta_p$, we approximate $\eta_p$ by the value of the total zero-shear viscosity. This yields the following estimate:
\begin{equation}
14<\frac{1}{1-a}<58.
\label{estimate}
\end{equation}

As  mentioned above, the exact value of the boundary between stable and unstable regions of the stability diagram proposed by Fielding \cite{Fielding:2010} depends on $\eta_s/\eta_p$ which is difficult to infer experimentally. Therefore it is possible that the lowest end of the range (\ref{estimate}) belongs to the region of the interfacial instability, while most of the range is in the stable region of the stability diagram.
Nevertheless, both are incompatible with the observation by Fardin \emph{et al.} \cite{Fardin:2009} of a bulk instability. 

Moreover, the model parameter $a=0.985$ used in this study 
\footnote{Note that the range (3) is based on the value of the relaxation time measured from linear rheology, which is somewhat different from the value that we found
by fitting the overal shape of the $\Sigma$ and $N_1$ curves. That is why our value of $\nicefrac{1}{(1-a)}$ lies slightly outside the range (\ref{estimate}).} gives $\nicefrac{1}{(1-a)}\approx 67$ --
well inside the stable region of the stability diagram of Fielding. We confirmed this by studying the linear stability with respect to the axisymmetric modes and found no instability. However, the flow \emph{does} exhibit a non-axisymmetric instability, which is the main finding of our paper.

\subsection{Schematic view of the original and auxiliary
  Taylor-Couette setups}

\begin{figure}[h]
\includegraphics[width=7.0cm, trim = 4cm 5cm 4cm 5cm, clip]{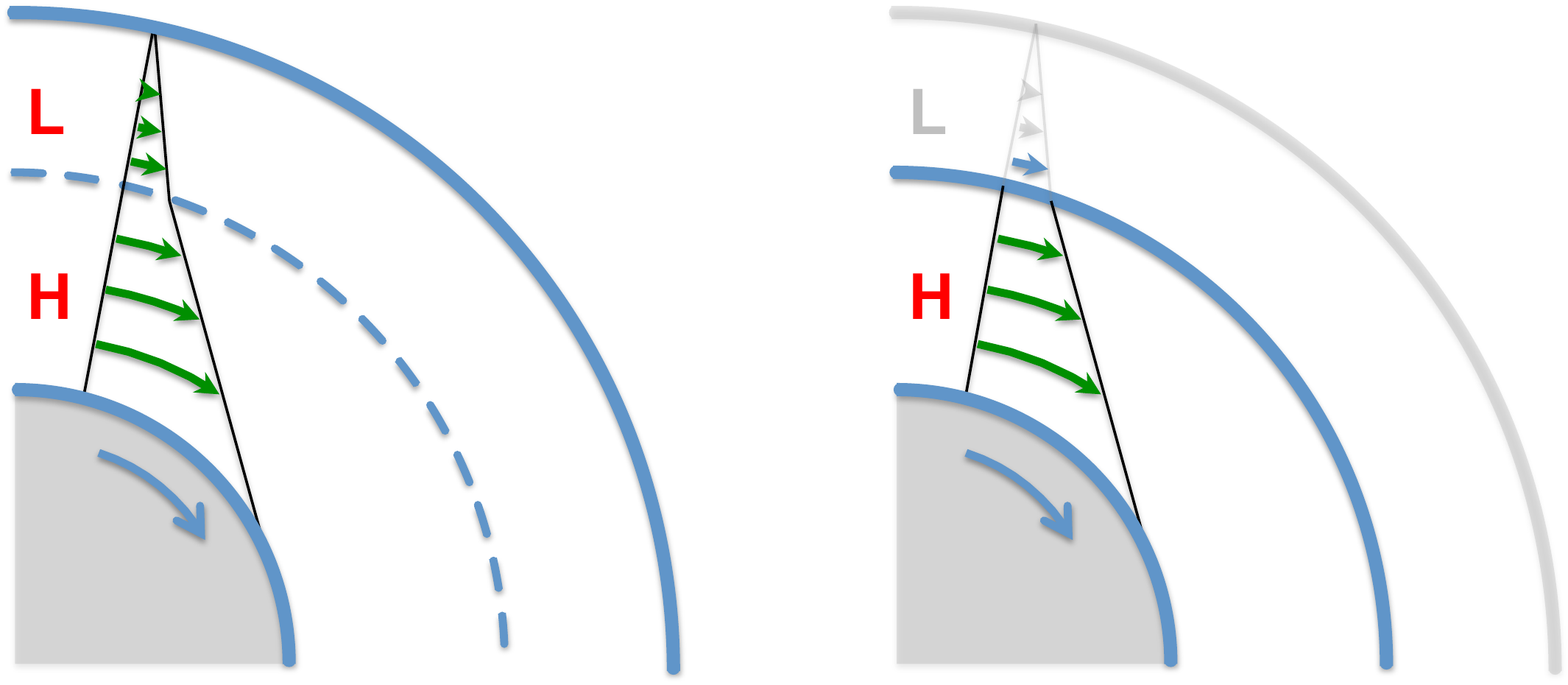}
\caption{The original (left) and auxiliary (right) Taylor-Couette
  setups. ``L'' and ``H'' denote the low- and high-shear-rate bands,
  respectively.}
\label{setups}
\end{figure}

In the auxiliary setup, the interface between the bands
  is replaced by a solid boundary that moves with the same velocity as
the low-shear-rate band in the original setup, see Fig.\ref{setups}. The only difference
between the two setups is the nature of the outer boundary
condition.  In the
original case, the velocities at the outer boundary of the
high-shear-rate band have to match the corresponding velocities in the
low-shear-rate band (a soft boundary), while in the auxiliary setup,
the presence of the rotating outer wall imposes the no-slip boundary
condition on the velocity.

At high Weissenberg numbers, the low-shear-rate band occupies only a
small portion of the gap in the original setup, and the
high-shear-rate band experiences the influence of the no-slip
condition imposed at the outer wall; the interface, therefore, acts as a
relatively hard boundary. In this case, the bulk eigenmodes in 
the spectra of the original and auxiliary systems coincide, as can be
seen from Fig.3(a). When the Weissenberg number is decreased,
corresponding to a wider low-shear-rate band, the boundary condition
at the interface starts to be ``softer'' and the agreement between the
eigenspectra of the original and auxiliary systems starts to
deteriorate.

\subsection{Velocity profiles associated with the interfacial and bulk modes at $Wi=6.4$ (see Fig.3(e))}

\hspace{1mm}

\begin{figure}[h]
\subfigure{\includegraphics[width=6.2cm]{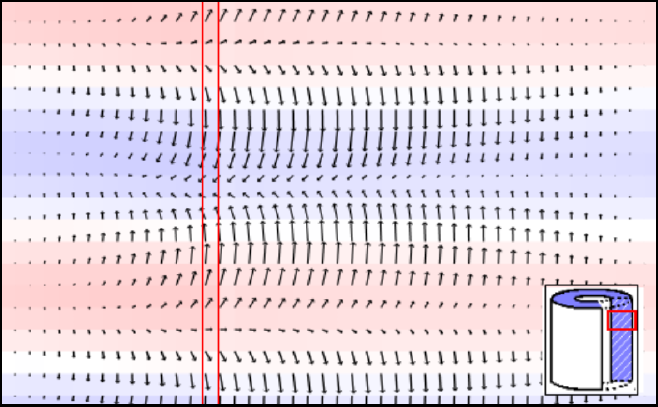}}
\subfigure{\includegraphics[width=6.2cm]{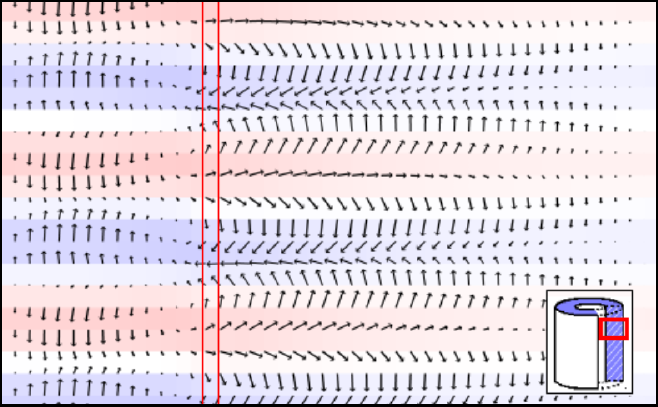}}
\caption{Velocity profiles in the $r-z$ plane for $Wi=6.4$ and
$l=4\,\mu m$:
(top) first peak of the dispersion relation at $k=14.4$ and $m=6$,
the interfacial mode; (bottom) second peak at $k=25$ and $m=24$, the
bulk mode. The red lines indicate the position and the width of the
 computational domain comprising the interface between the bands.
}
\end{figure}

\bibliography{bibshearbanding}

\begin{thebibliography}{34}%
\makeatletter
\providecommand \@ifxundefined [1]{%
 \@ifx{#1\undefined}
}%
\providecommand \@ifnum [1]{%
 \ifnum #1\expandafter \@firstoftwo
 \else \expandafter \@secondoftwo
 \fi
}%
\providecommand \@ifx [1]{%
 \ifx #1\expandafter \@firstoftwo
 \else \expandafter \@secondoftwo
 \fi
}%
\providecommand \natexlab [1]{#1}%
\providecommand \enquote  [1]{``#1''}%
\providecommand \bibnamefont  [1]{#1}%
\providecommand \bibfnamefont [1]{#1}%
\providecommand \citenamefont [1]{#1}%
\providecommand \href@noop [0]{\@secondoftwo}%
\providecommand \href [0]{\begingroup \@sanitize@url \@href}%
\providecommand \@href[1]{\@@startlink{#1}\@@href}%
\providecommand \@@href[1]{\endgroup#1\@@endlink}%
\providecommand \@sanitize@url [0]{\catcode `\\12\catcode `\$12\catcode
  `\&12\catcode `\#12\catcode `\^12\catcode `\_12\catcode `\%12\relax}%
\providecommand \@@startlink[1]{}%
\providecommand \@@endlink[0]{}%
\providecommand \url  [0]{\begingroup\@sanitize@url \@url }%
\providecommand \@url [1]{\endgroup\@href {#1}{\urlprefix }}%
\providecommand \urlprefix  [0]{URL }%
\providecommand \Eprint [0]{\href }%
\providecommand \doibase [0]{http://dx.doi.org/}%
\providecommand \selectlanguage [0]{\@gobble}%
\providecommand \bibinfo  [0]{\@secondoftwo}%
\providecommand \bibfield  [0]{\@secondoftwo}%
\providecommand \translation [1]{[#1]}%
\providecommand \BibitemOpen [0]{}%
\providecommand \bibitemStop [0]{}%
\providecommand \bibitemNoStop [0]{.\EOS\space}%
\providecommand \EOS [0]{\spacefactor3000\relax}%
\providecommand \BibitemShut  [1]{\csname bibitem#1\endcsname}%
\let\auto@bib@innerbib\@empty
\bibitem [{\citenamefont {Larson}(1999)}]{Larson:book}%
  \BibitemOpen
  \bibfield  {author} {\bibinfo {author} {\bibfnamefont {R.~G.}\ \bibnamefont
  {Larson}},\ }\href@noop {} {\emph {\bibinfo {title} {The Structure and
  Rheology of Complex Fluids}}}\ (\bibinfo  {publisher} {Oxford University
  Press},\ \bibinfo {year} {1999})\BibitemShut {NoStop}%
\bibitem [{\citenamefont {Cates}\ and\ \citenamefont
  {Fielding}(2006)}]{Cates:2006}%
  \BibitemOpen
  \bibfield  {author} {\bibinfo {author} {\bibfnamefont {M.~E.}\ \bibnamefont
  {Cates}}\ and\ \bibinfo {author} {\bibfnamefont {S.~M.}\ \bibnamefont
  {Fielding}},\ }\href@noop {} {\bibfield  {journal} {\bibinfo  {journal} {Adv.
  in Phys.}\ }\textbf {\bibinfo {volume} {55}},\ \bibinfo {pages} {799}
  (\bibinfo {year} {2006})}\BibitemShut {NoStop}%
\bibitem [{\citenamefont {Lerouge}\ and\ \citenamefont
  {Berret}(2010)}]{Lerouge:2010}%
  \BibitemOpen
  \bibfield  {author} {\bibinfo {author} {\bibfnamefont {S.}~\bibnamefont
  {Lerouge}}\ and\ \bibinfo {author} {\bibfnamefont {J.-F.}\ \bibnamefont
  {Berret}},\ }\href@noop {} {\bibfield  {journal} {\bibinfo  {journal} {Adv
  Polym Sci}\ }\textbf {\bibinfo {volume} {230}},\ \bibinfo {pages} {1}
  (\bibinfo {year} {2010})}\BibitemShut {NoStop}%
\bibitem [{\citenamefont {Schall}\ and\ \citenamefont {van
  Hecke}(2010)}]{Hecke:2010}%
  \BibitemOpen
  \bibfield  {author} {\bibinfo {author} {\bibfnamefont {P.}~\bibnamefont
  {Schall}}\ and\ \bibinfo {author} {\bibfnamefont {M.}~\bibnamefont {van
  Hecke}},\ }\href@noop {} {\bibfield  {journal} {\bibinfo  {journal} {Annu.
  Rev. Fluid Mech.}\ }\textbf {\bibinfo {volume} {42}},\ \bibinfo {pages} {67}
  (\bibinfo {year} {2010})}\BibitemShut {NoStop}%
\bibitem [{\citenamefont {{Gilbreth \emph{et al.}}}(2006)}]{Gilbreth:2006}%
  \BibitemOpen
  \bibfield  {author} {\bibinfo {author} {\bibfnamefont {C.}~\bibnamefont
  {{Gilbreth \emph{et al.}}}},\ }\href@noop {} {\bibfield  {journal} {\bibinfo
  {journal} {Phys. Rev. E}\ }\textbf {\bibinfo {volume} {74}},\ \bibinfo
  {pages} {051406} (\bibinfo {year} {2006})}\BibitemShut {NoStop}%
\bibitem [{\citenamefont {{Chen \emph{et al.}}}(1992)}]{Chen:1992}%
  \BibitemOpen
  \bibfield  {author} {\bibinfo {author} {\bibfnamefont {L.~B.}\ \bibnamefont
  {{Chen \emph{et al.}}}},\ }\href@noop {} {\bibfield  {journal} {\bibinfo
  {journal} {Phys. Rev. Lett.}\ }\textbf {\bibinfo {volume} {69}},\ \bibinfo
  {pages} {688} (\bibinfo {year} {1992})}\BibitemShut {NoStop}%
\bibitem [{\citenamefont {{Besseling \emph{et al.}}}(2010)}]{Rut:2010}%
  \BibitemOpen
  \bibfield  {author} {\bibinfo {author} {\bibfnamefont {R.}~\bibnamefont
  {{Besseling \emph{et al.}}}},\ }\href@noop {} {\bibfield  {journal} {\bibinfo
   {journal} {Phys. Rev. Lett.}\ }\textbf {\bibinfo {volume} {105}},\ \bibinfo
  {pages} {268301} (\bibinfo {year} {2010})}\BibitemShut {NoStop}%
\bibitem [{\citenamefont {Rehage}\ and\ \citenamefont
  {Hoffmann}(1991)}]{Rehage:1991}%
  \BibitemOpen
  \bibfield  {author} {\bibinfo {author} {\bibfnamefont {H.}~\bibnamefont
  {Rehage}}\ and\ \bibinfo {author} {\bibfnamefont {H.}~\bibnamefont
  {Hoffmann}},\ }\href@noop {} {\bibfield  {journal} {\bibinfo  {journal} {Mol.
  Phys.}\ }\textbf {\bibinfo {volume} {74}},\ \bibinfo {pages} {933} (\bibinfo
  {year} {1991})}\BibitemShut {NoStop}%
\bibitem [{\citenamefont {Cates}(1996)}]{Cates:1996}%
  \BibitemOpen
  \bibfield  {author} {\bibinfo {author} {\bibfnamefont {M.~E.}\ \bibnamefont
  {Cates}},\ }\href@noop {} {\bibfield  {journal} {\bibinfo  {journal} {J.
  Phys.: Condens. Matter}\ }\textbf {\bibinfo {volume} {8}},\ \bibinfo {pages}
  {9167} (\bibinfo {year} {1996})}\BibitemShut {NoStop}%
\bibitem [{\citenamefont {{Lerouge \emph{et al.}}}(2006)}]{Lerouge:2006}%
  \BibitemOpen
  \bibfield  {author} {\bibinfo {author} {\bibfnamefont {S.}~\bibnamefont
  {{Lerouge \emph{et al.}}}},\ }\href@noop {} {\bibfield  {journal} {\bibinfo
  {journal} {Phys. Rev. Lett.}\ }\textbf {\bibinfo {volume} {96}},\ \bibinfo
  {pages} {088301} (\bibinfo {year} {2006})}\BibitemShut {NoStop}%
\bibitem [{\citenamefont {{Becu \emph{et al.}}}(2007)}]{Becu:2007}%
  \BibitemOpen
  \bibfield  {author} {\bibinfo {author} {\bibfnamefont {L.}~\bibnamefont
  {{Becu \emph{et al.}}}},\ }\href@noop {} {\bibfield  {journal} {\bibinfo
  {journal} {Phys. Rev. E}\ }\textbf {\bibinfo {volume} {76}},\ \bibinfo
  {pages} {011503} (\bibinfo {year} {2007})}\BibitemShut {NoStop}%
\bibitem [{\citenamefont {{Lerouge \emph{et al.}}}(2008)}]{Lerouge:2008}%
  \BibitemOpen
  \bibfield  {author} {\bibinfo {author} {\bibfnamefont {S.}~\bibnamefont
  {{Lerouge \emph{et al.}}}},\ }\href@noop {} {\bibfield  {journal} {\bibinfo
  {journal} {Soft Matter}\ }\textbf {\bibinfo {volume} {4}},\ \bibinfo {pages}
  {1808} (\bibinfo {year} {2008})}\BibitemShut {NoStop}%
\bibitem [{\citenamefont {{Fardin \emph{et al.}}}(2009)}]{Fardin:2009}%
  \BibitemOpen
  \bibfield  {author} {\bibinfo {author} {\bibfnamefont {M.~A.}\ \bibnamefont
  {{Fardin \emph{et al.}}}},\ }\href@noop {} {\bibfield  {journal} {\bibinfo
  {journal} {Phys. Rev. Lett.}\ }\textbf {\bibinfo {volume} {103}},\ \bibinfo
  {pages} {028302} (\bibinfo {year} {2009})}\BibitemShut {NoStop}%
\bibitem [{\citenamefont {Renardy}(1988)}]{Renardy:1988}%
  \BibitemOpen
  \bibfield  {author} {\bibinfo {author} {\bibfnamefont {Y.}~\bibnamefont
  {Renardy}},\ }\href@noop {} {\bibfield  {journal} {\bibinfo  {journal} {J. of
  Non-Newtonian Fluid Mech.}\ }\textbf {\bibinfo {volume} {28}},\ \bibinfo
  {pages} {99} (\bibinfo {year} {1988})}\BibitemShut {NoStop}%
\bibitem [{\citenamefont {{Hinch \emph{et al.}}}(1992)}]{Hinch:1992}%
  \BibitemOpen
  \bibfield  {author} {\bibinfo {author} {\bibfnamefont {E.~J.}\ \bibnamefont
  {{Hinch \emph{et al.}}}},\ }\href@noop {} {\bibfield  {journal} {\bibinfo
  {journal} {J. of Non-Newtonian Fluid Mech.}\ }\textbf {\bibinfo {volume}
  {43}},\ \bibinfo {pages} {311} (\bibinfo {year} {1992})}\BibitemShut
  {NoStop}%
\bibitem [{\citenamefont {Renardy}\ and\ \citenamefont
  {Renardy}(1999)}]{Renardy:1999}%
  \BibitemOpen
  \bibfield  {author} {\bibinfo {author} {\bibfnamefont {Y.}~\bibnamefont
  {Renardy}}\ and\ \bibinfo {author} {\bibfnamefont {M.}~\bibnamefont
  {Renardy}},\ }\href@noop {} {\bibfield  {journal} {\bibinfo  {journal} {J. of
  Non-Newtonian Fluid Mech.}\ }\textbf {\bibinfo {volume} {81}},\ \bibinfo
  {pages} {215} (\bibinfo {year} {1999})}\BibitemShut {NoStop}%
\bibitem [{\citenamefont {{Larson \emph{et al.}}}(1990)}]{Larson:1990}%
  \BibitemOpen
  \bibfield  {author} {\bibinfo {author} {\bibfnamefont {R.~G.}\ \bibnamefont
  {{Larson \emph{et al.}}}},\ }\href@noop {} {\bibfield  {journal} {\bibinfo
  {journal} {J. Fluid Mech.}\ }\textbf {\bibinfo {volume} {218}},\ \bibinfo
  {pages} {573} (\bibinfo {year} {1990})}\BibitemShut {NoStop}%
\bibitem [{\citenamefont {Pakdel}\ and\ \citenamefont
  {McKinley}(1996)}]{Pakdel:1996}%
  \BibitemOpen
  \bibfield  {author} {\bibinfo {author} {\bibfnamefont {P.}~\bibnamefont
  {Pakdel}}\ and\ \bibinfo {author} {\bibfnamefont {G.~H.}\ \bibnamefont
  {McKinley}},\ }\href@noop {} {\bibfield  {journal} {\bibinfo  {journal}
  {Phys. Rev. Lett.}\ }\textbf {\bibinfo {volume} {77}},\ \bibinfo {pages}
  {2459} (\bibinfo {year} {1996})}\BibitemShut {NoStop}%
\bibitem [{\citenamefont {Fielding}(2010)}]{Fielding:2010}%
  \BibitemOpen
  \bibfield  {author} {\bibinfo {author} {\bibfnamefont {S.~M.}\ \bibnamefont
  {Fielding}},\ }\href@noop {} {\bibfield  {journal} {\bibinfo  {journal}
  {Phys. Rev. Lett.}\ }\textbf {\bibinfo {volume} {104}},\ \bibinfo {pages}
  {198303} (\bibinfo {year} {2010})}\BibitemShut {NoStop}%
\bibitem [{\citenamefont {Fielding}(2005)}]{Fielding:2005}%
  \BibitemOpen
  \bibfield  {author} {\bibinfo {author} {\bibfnamefont {S.~M.}\ \bibnamefont
  {Fielding}},\ }\href@noop {} {\bibfield  {journal} {\bibinfo  {journal}
  {Phys. Rev. Lett.}\ }\textbf {\bibinfo {volume} {95}},\ \bibinfo {pages}
  {134501} (\bibinfo {year} {2005})}\BibitemShut {NoStop}%
\bibitem [{\citenamefont {Fielding}\ and\ \citenamefont
  {Olmsted}(2006)}]{Fielding:2006}%
  \BibitemOpen
  \bibfield  {author} {\bibinfo {author} {\bibfnamefont {S.~M.}\ \bibnamefont
  {Fielding}}\ and\ \bibinfo {author} {\bibfnamefont {P.~D.}\ \bibnamefont
  {Olmsted}},\ }\href@noop {} {\bibfield  {journal} {\bibinfo  {journal} {Phys.
  Rev. Lett.}\ }\textbf {\bibinfo {volume} {96}},\ \bibinfo {pages} {104502}
  (\bibinfo {year} {2006})}\BibitemShut {NoStop}%
\bibitem [{\citenamefont {Wilson}\ and\ \citenamefont
  {Fielding}(2006)}]{Wilson:2006}%
  \BibitemOpen
  \bibfield  {author} {\bibinfo {author} {\bibfnamefont {H.~J.}\ \bibnamefont
  {Wilson}}\ and\ \bibinfo {author} {\bibfnamefont {S.~M.}\ \bibnamefont
  {Fielding}},\ }\href@noop {} {\bibfield  {journal} {\bibinfo  {journal} {J.
  of Non-Newtonian Fluid Mech.}\ }\textbf {\bibinfo {volume} {138}},\ \bibinfo
  {pages} {181} (\bibinfo {year} {2006})}\BibitemShut {NoStop}%
\bibitem [{\citenamefont {Fielding}(2007)}]{Fielding:2007a}%
  \BibitemOpen
  \bibfield  {author} {\bibinfo {author} {\bibfnamefont {S.~M.}\ \bibnamefont
  {Fielding}},\ }\href@noop {} {\bibfield  {journal} {\bibinfo  {journal}
  {Phys. Rev. E}\ }\textbf {\bibinfo {volume} {76}},\ \bibinfo {pages} {016311}
  (\bibinfo {year} {2007})}\BibitemShut {NoStop}%
\bibitem [{\citenamefont {Johnson}\ and\ \citenamefont
  {Segalman}(1977)}]{Johnson:1977}%
  \BibitemOpen
  \bibfield  {author} {\bibinfo {author} {\bibfnamefont {M.~W.}\ \bibnamefont
  {Johnson}}\ and\ \bibinfo {author} {\bibfnamefont {D.}~\bibnamefont
  {Segalman}},\ }\href@noop {} {\bibfield  {journal} {\bibinfo  {journal} {J.
  of Non-Newtonian Fluid Mech.}\ }\textbf {\bibinfo {volume} {2}},\ \bibinfo
  {pages} {255} (\bibinfo {year} {1977})}\BibitemShut {NoStop}%
\bibitem [{\citenamefont {{Olmsted \emph{et al.}}}(2000)}]{Olmsted:2000}%
  \BibitemOpen
  \bibfield  {author} {\bibinfo {author} {\bibfnamefont {P.~D.}\ \bibnamefont
  {{Olmsted \emph{et al.}}}},\ }\href@noop {} {\bibfield  {journal} {\bibinfo
  {journal} {J. Rheol.}\ }\textbf {\bibinfo {volume} {44}},\ \bibinfo {pages}
  {257} (\bibinfo {year} {2000})}\BibitemShut {NoStop}%
\bibitem [{\citenamefont {{Ballesta \emph{et al.}}}(2007)}]{Ballesta:2007}%
  \BibitemOpen
  \bibfield  {author} {\bibinfo {author} {\bibfnamefont {P.}~\bibnamefont
  {{Ballesta \emph{et al.}}}},\ }\href@noop {} {\bibfield  {journal} {\bibinfo
  {journal} {J. Rheol.}\ }\textbf {\bibinfo {volume} {51}},\ \bibinfo {pages}
  {1047} (\bibinfo {year} {2007})}\BibitemShut {NoStop}%
\bibitem [{\citenamefont {{Masselon \emph{et al.}}}(2008)}]{Masselon:2008}%
  \BibitemOpen
  \bibfield  {author} {\bibinfo {author} {\bibfnamefont {C.}~\bibnamefont
  {{Masselon \emph{et al.}}}},\ }\href@noop {} {\bibfield  {journal} {\bibinfo
  {journal} {Phys. Rev. Lett.}\ }\textbf {\bibinfo {volume} {100}},\ \bibinfo
  {pages} {038301} (\bibinfo {year} {2008})}\BibitemShut {NoStop}%
\bibitem [{\citenamefont {{Fardin \emph{et al.}}}(2010)}]{Fardin:2010}%
  \BibitemOpen
  \bibfield  {author} {\bibinfo {author} {\bibfnamefont {M.~A.}\ \bibnamefont
  {{Fardin \emph{et al.}}}},\ }\href@noop {} {\bibfield  {journal} {\bibinfo
  {journal} {Phys. Rev. Lett.}\ }\textbf {\bibinfo {volume} {104}},\ \bibinfo
  {pages} {178303} (\bibinfo {year} {2010})}\BibitemShut {NoStop}%
\bibitem [{\citenamefont {Canuto}\ \emph {et~al.}(1988)\citenamefont {Canuto},
  \citenamefont {Hussaini}, \citenamefont {Quarteroni},\ and\ \citenamefont
  {Zang}}]{Canuto:book}%
  \BibitemOpen
  \bibfield  {author} {\bibinfo {author} {\bibfnamefont {C.}~\bibnamefont
  {Canuto}}, \bibinfo {author} {\bibfnamefont {M.}~\bibnamefont {Hussaini}},
  \bibinfo {author} {\bibfnamefont {A.}~\bibnamefont {Quarteroni}}, \ and\
  \bibinfo {author} {\bibfnamefont {T.}~\bibnamefont {Zang}},\ }\href@noop {}
  {\emph {\bibinfo {title} {Spectral Methods in Fluid Dynamics}}}\ (\bibinfo
  {publisher} {Springer Verlag},\ \bibinfo {year} {1988})\BibitemShut {NoStop}%
\bibitem [{\citenamefont {{Cromer \emph{et al.}}}(2011)}]{Cromer:2011}%
  \BibitemOpen
  \bibfield  {author} {\bibinfo {author} {\bibfnamefont {M.}~\bibnamefont
  {{Cromer \emph{et al.}}}},\ }\href@noop {} {\bibfield  {journal} {\bibinfo
  {journal} {J. of Non-Newtonian Fluid Mech.}\ }\textbf {\bibinfo {volume}
  {166}},\ \bibinfo {pages} {566} (\bibinfo {year} {2011})}\BibitemShut
  {NoStop}%
\bibitem [{\citenamefont {{Fardin \emph{et al.}}}(2011)}]{Fardin:2011}%
  \BibitemOpen
  \bibfield  {author} {\bibinfo {author} {\bibfnamefont {M.~A.}\ \bibnamefont
  {{Fardin \emph{et al.}}}},\ }\href@noop {} {\  (\bibinfo {year} {2011})},\
  \Eprint {http://arxiv.org/abs/cond-matt/1108.2731}
  {arXiv:cond-matt/1108.2731} \BibitemShut {NoStop}%
\bibitem [{\citenamefont {Avgousti}\ and\ \citenamefont
  {Beris}(1993)}]{Avgousti:1993}%
  \BibitemOpen
  \bibfield  {author} {\bibinfo {author} {\bibfnamefont {M.}~\bibnamefont
  {Avgousti}}\ and\ \bibinfo {author} {\bibfnamefont {A.~N.}\ \bibnamefont
  {Beris}},\ }\href@noop {} {\bibfield  {journal} {\bibinfo  {journal} {J. of
  Non-Newtonian Fluid Mech.}\ }\textbf {\bibinfo {volume} {50}},\ \bibinfo
  {pages} {225} (\bibinfo {year} {1993})}\BibitemShut {NoStop}%
\bibitem [{\citenamefont {{Decruppe \emph{et al.}}}(2010)}]{Decruppe:2010}%
  \BibitemOpen
  \bibfield  {author} {\bibinfo {author} {\bibfnamefont {J.~P.}\ \bibnamefont
  {{Decruppe \emph{et al.}}}},\ }\href@noop {} {\bibfield  {journal} {\bibinfo
  {journal} {Phys. Rev. Lett.}\ }\textbf {\bibinfo {volume} {105}},\ \bibinfo
  {pages} {258301} (\bibinfo {year} {2010})}\BibitemShut {NoStop}%
\bibitem [{Note1()}]{Note1}%
  \BibitemOpen
  \bibinfo {note} {Note that the range (3) is based on the value of the
  relaxation time measured from linear rheology, which is somewhat different
  from the value that we found by fitting the overal shape of the $\Sigma $ and
  $N_1$ curves. That is why our value of $\protect \nicefrac {1}{(1-a)}$ lies
  slightly outside the range (\ref {estimate}).}\BibitemShut {Stop}%
\end{thebibliography}%
\end{document}